\documentclass[fleqn,usenatbib]{mnras}

\usepackage[T1]{fontenc}
\DeclareRobustCommand{\VAN}[3]{#2}
\let\VANthebibliography\thebibliography
\def\thebibliography{\DeclareRobustCommand{\VAN}[3]{##3}\VANthebibliography}
\usepackage{graphicx}
\usepackage{amsmath, amssymb}
\usepackage{threeparttable,multirow}
\usepackage{newtxtext,newtxmath}
\usepackage{color,soul}
\usepackage{orcidlink}

\def\kms{\,km\,s$^{-1}$}
\def\esig{$\sigma_{\rm e}$}
\def\msig{$M_*$--$\sigma_{\rm e}$}

\def\re{$\,R_{\rm e}$}

\title[Kinematic scaling of thin and thick discs]{Kinematic scaling of thin and thick discs from SAMI to NewHorizon}
\author[S.~Oh et~al.]{Sree~Oh\orcidlink{0000-0002-4731-9604},$^{1}$\thanks{E-mail: sreemario@gmail.com} 
J. K.~Jang\orcidlink{0000-0002-0858-5264},$^{1}$
Giulia~Santucci\orcidlink{0000-0003-3283-4686},$^{2}$
Sukyoung K. Yi\orcidlink{0000-0002-4556-2619},$^{1}$
Matthew~Colless\orcidlink{0000-0001-9552-8075},$^{3}$
Scott M. Croom\orcidlink{0000-0003-2880-9197},$^{4}$
\newauthor
Yohan Dubois\orcidlink{0000-0003-0225-6387},$^{5}$
Stefania Barsanti\orcidlink{0000-0002-9332-5386},$^{4}$
Christophe Pichon\orcidlink{0000-0003-0695-6735},$^{5,6}$
S\'ebastien Peirani\orcidlink{0000-0001-6902-2898},$^{7,8,5}$
\newauthor
Madusha L. P. Gunawardhana\orcidlink{0000-0002-7301-461X}, $^{4,3,9}$
\\ ~ \\
$^{1}$Department of Astronomy and Yonsei University Observatory, Yonsei University, Seoul, 03722, Republic of Korea\\
$^{2}$CSIRO Space \& Astronomy, PO Box 1130, Bentley, Western Australia 6102, Australia\\
$^{3}$Research School of Astronomy and Astrophysics, Australian National University, Canberra, ACT 2611, Australia\\
$^{4}$Sydney Institute for Astronomy (SIfA), School of Physics, The University of Sydney, NSW 2006, Australia\\
$^{5}$Institut d’Astrophysique de Paris, CNRS and Sorbonne Université, UMR 7095, 98 bis Boulevard Arago, F-75014 Paris, France\\
$^{6}$Kyung Hee University, Dept. of Astronomy \& Space Science, Yongin-shi, Gyeonggi-do 17104, Republic of Korea\\
$^{7}$ILANCE, CNRS – University of Tokyo International Research Laboratory, Kashiwa, Chiba 277-8582, Japan\\
$^{8}$Kavli IPMU (WPI), UTIAS, The University of Tokyo, Kashiwa, Chiba 277-8583, Japan\\
$^{9}$ARC Centre of Excellence for All Sky Astrophysics in 3 Dimensions (ASTRO 3D), Australia\\
}
\date{Accepted XXX. Received YYY; in original form ZZZ}

\begin{document}
\label{firstpage}
\pagerange{\pageref{firstpage}--\pageref{lastpage}}\pubyear{2021}
\maketitle

\begin{abstract}
We revisit the relation between disc stellar mass and disc velocity dispersion (\msig) and extend it to thin and thick subcomponents using orbit-based dynamical models of 161 SAMI galaxies and counterpart measurements for 31 disc galaxies in the NewHorizon simulation. On the observational side, we apply Schwarzschild orbit superposition to recover orbital circularity distributions and component kinematics. On the simulation side, we sample thin and thick discs by circularity and, separately, by stellar age to test classification dependence. Our analysis reveals three main results. (1) Discs follow a tight \msig\ relation, nearly parallel to the bulge relation. (2) For both circularity- and age-based definitions, the thick-disc component is systematically hotter than the thin-disc component, and the thin-thick dispersion ratio varies only weakly with mass. However, age cuts yield a smaller kinematic contrast, indicating that stellar age and orbital circularity do not map one-to-one and that no single global age threshold reproduces the circularity-based split. (3) Method and data systematics are present, with Schwarzschild modelling returning slightly higher disc \esig\ than spectroscopic bulge--disc decompositions, and simulated discs showing lower \esig\ at fixed mass than observed. All these results are consistent with a baseline set by vertical-equilibrium scalings, with secular heating accumulating over time and modulating the dispersion at fixed mass. Occasional minor interactions may add localised heating but do not appear to be essential for explaining the qualitative, global trends reported here. Future tests with chemo-dynamical modelling and higher-resolution, chemistry-tracking simulations will provide stronger constraints on disc substructures in external galaxies.
\end{abstract}


\section{Introduction}
\label{sec:intro}

Disc galaxies are predominantly rotation-supported systems, yet recent studies decomposing the kinematics into individual components show that their disc components obey a well-defined stellar mass–velocity dispersion (\msig) relation, with slope and scatter broadly comparable to those of bulges or ellipticals (Oh et al.\ 2020; Irodotou \& Thomas 2021). This highlights the physical relevance of stellar velocity dispersion in discs alongside traditional rotation-based scalings (i.e.,\ the Tully-Fisher relation). 

Decades of studies have established that galaxy discs commonly comprise thin and thick components (e.g.,\ Gilmore \& Reid 1983; Dalcanton \& Bernstein 2002; Comer\'on et al.\ 2011). Traditionally, these components were identified from the vertical light distribution. Subsequent studies on the Milky Way broadened this picture by incorporating differences in chemical abundances, stellar ages, and kinematics, revealing a more complex, multi-dimensional separation between the two (e.g.,\ Yoachim \& Dalcanton 2008; Bovy et al.\ 2012; Haywood et al.\ 2013; Bensby et al.\ 2014). {In the Milky Way, the thick disc is typically older and more $\alpha$-enhanced than the thin disc, but its inferred structure depends on how the two components are defined. In particular, chemically defined high-$\alpha$ populations tend to have a shorter radial scale length than geometrically defined thick-disc selections, and this difference may be related to the flaring of mono-abundance populations (Minchev et al. 2017).} These complexities motivate a component-resolved kinematic analysis for disc substructures and their scaling relations.

Measuring kinematics in discs and their subcomponents is non-trivial: line-of-sight projection mixes anisotropic motions, and inclination, seeing, and template/weighting choices can bias dispersion estimates. Spectroscopic decomposition of integral-field spectroscopy (IFS) data enables simultaneous fitting of multiple kinematic components, but in practice it has mostly been used to separate bulge and disc (Tabor et al.\ 2019; Oh et al.\ 2020; Pak et al.\ 2021). Extending this approach to thin and thick discs remains difficult because current wide-field IFS surveys often lack the signal-to-noise and spatial resolution needed to break degeneracies, and PSF smearing further blends the components. {Surveys using higher-resolution instruments, such as GECKOS (van de Sande et al.\ 2024), illustrate potential strategies to overcome these challenges.}

Orbit-based dynamical modelling now makes component-resolved kinematics accessible from IFS data. In particular, Schwarzschild orbit superposition recovers the orbital distribution from two-dimensional (2D) maps and separates cold, warm, and hot components by orbital circularity, measuring kinematics for each component (e.g.,\ Schwarzschild 1979; van den Bosch et al.\ 2008; Zhu et al.\ 2018; Thater et al.\ 2022). In parallel, high-resolution cosmological simulations and zoom-ins provide particle orbits and ages, allowing thin and thick disc sampling and direct measure of their particle kinematics (Park et al.\ 2021; Yi et al.\ 2024). 

Building on these developments, we revisit the disc \msig\ relation and extend it to thin and thick subcomponents. {We measure \esig\ as the flux-weighted mean within 1\re, using 2D velocity dispersion maps} from the Sydney-AAO Multi-object Integral field spectroscopy (SAMI) Galaxy Survey (Croom et al.\ 2012; Bryant et al.\ 2015) and galaxies from the NewHorizon (NH; Dubois et al.\ 2021) simulation. On the observational side, {we use the data from Santucci et al.\ (2022),} who fit Schwarzschild models to the SAMI data to recover orbital circularity distributions and component kinematics. On the simulation side, we define thin and thick discs by circularity and, separately, by particle age to test how the classification choice affects the inferred trends. In this paper we present the disc \msig\ relation in SAMI and NH and compare it with bulges, quantify the thin–thick dispersion contrast under different sampling criteria, assess systematic differences between Schwarzschild and spectroscopic decomposition and between observation and simulation, and discuss the physical implications. 

The paper is organised as follows. Section~2 describes the data and samples (SAMI and NH). Section~3 details the component sampling based on orbital circularity and stellar age. Section~4 presents the \msig\ relations for discs and for thin/thick components. Section~5 discusses method and observation–simulation systematics, the physical origin of the disc \msig\ relation, and implications for the origin of thin and thick discs. Section~6 summarises our conclusions. 
 
\section{Data and Sample}
\label{sec:data}

\subsection{The SAMI Galaxy Survey}
\label{sec:sami}

The final and third data release of the SAMI Galaxy Survey includes more than 3000 unique galaxy cubes at redshifts $z < 0.095$ (Bryant et~al.\ 2015; Croom et~al.\ 2021). Galaxies were targeted within a series of stepped volumes in the redshift and mass plane (see Fig. 4 of Bryant et al.\ 2015), where a stellar mass has been derived using $i$-band photometry and $g-i$ colour (Taylor et al.\ 2011). The majority of SAMI targets are drawn from the three equatorial regions of the Galaxy And Mass Assembly (GAMA; Driver et~al.\ 2011) survey. In addition, eight cluster regions have been included to complete the environmental matrix (Owers et~al.\ 2017). 

The SAMI instrument is a multi-object fibre integral field system (Croom et~al.\ 2012), composed of 13 hexabundles, each containing 61 1.6 arcsec-diameter optical fibres (Bland-Hawthorn et~al.\ 2011; Bryant et~al.\ 2011, 2014). The hexabundles are fed into the AAOmega dual-arm spectrograph mounted on the 3.9-metre Anglo-Australian Telescope (Sharp et~al.\ 2006). The 580V and 1000R gratings are used for blue (3750--5750\,\AA) and red (6300--7400\,\AA) arms, yielding the spectral resolution of R = 1808 and R = 4304 respectively. The spectral resolutions in the blue and red arms are equivalent to an effective velocity dispersion of 70.4 and 29.6\kms~respectively (van de Sande et~al.\ 2017b). The SAMI Galaxy Survey adopts a Lambda-cold dark matter ($\Lambda$CDM) cosmology with $\rm \Omega_m = 0.3, ~\Omega_{\Lambda}= 0.7,~and~H_0 = 70~km~s^{-1} Mpc^{-1}$.

\subsection{NewHorizon simulation}
We use the NH simulation, a high-resolution cosmological hydrodynamic simulation with adaptive mesh refinement (AMR), {\sc RAMSES} (Teyssier 2002). The simulation volume has extracted from its parent Horizon-AGN simulation (Dubois et al.\ 2016) with a box size of 142 Mpc and zoomed-in on a sphere with diameter 20 Mpc comoving targeting a field (average density) environment. The simulation has reached z = 0.17 with a high spatial resolution of 34 pc (Yi et al. 2024). The simulation adopts a Lambda-cold dark matter ($\Lambda$CDM) cosmology comparable with the WMAP-7 result (Komatsu et al.\ 2011) where $\rm \Omega_m = 0.272$, $\Omega_{\Lambda}= 0.728$,~$\sigma_8$=0.81,~$\rm \Omega_b = 0.045,~and~H_0 = 70.4~km~s^{-1} Mpc^{-1}$. 

The cooling of the primordial and metal-enriched gas (Sutherland \& Dopita 1993; Dalgarno \& McCray 1972) and the heating from a uniform ultraviolet background after the reionization at $z=10$ (Haardt \& Madau 1996) are modelled to consider radiative cooling and heating. Star formation occurs when a hydrogen number density of a cell is greater than $\rm n_H$ = 10 cm$^{-3}$ and the temperature is lower than 2 x 10$^4$ K, following the Schmidt law, but adopting varying star formation efficiency (Kimm et al.\ 2017; Trebitsch et al.\ 2017). The mechanical supernova feedback is modelled to return 31\% of mass from a stellar particle when it explodes at the age of 5 Myr (Kimm \& Cen 2014; Kimm et al.\ 2015), while assuming the stellar particles follow a Chabrier initial mass function (Chabrier 2005). Sink particles (i.e.\ black holes) are formed in cells where the gas and stellar densities are above the threshold for star formation and grow with the Bondi-Hoyle-Lyttleton accretion model (Hoyle \& Lyttleton 1939; Bondi \& Hoyle 1944). Active galactic nuclei feedback is applied either in the radio jet (Dubois et al.\ 2010) or in the quasar heating modes (Teyssier et al.\ 2011) depending on the Eddington ratio (Dubois et al.\ 2012). Galaxies, composed of at least 50 stellar particles {(each with a mass of $10^{4}$ $\rm M_{\odot}$)}, are identified together with dark matter halos using the AdaptaHOP halo finder (Aubert et al.\ 2004). At $z=0.17$, 366 galaxies with stellar mass greater than $10^{7}$ $\rm M_{\odot}$ are identified. See Dubois et al.\ (2021) for a detailed description of NH simulation. 

\subsection{Sample} \label{sec:sam}
\begin{figure}
\centering
\includegraphics[width=\columnwidth]{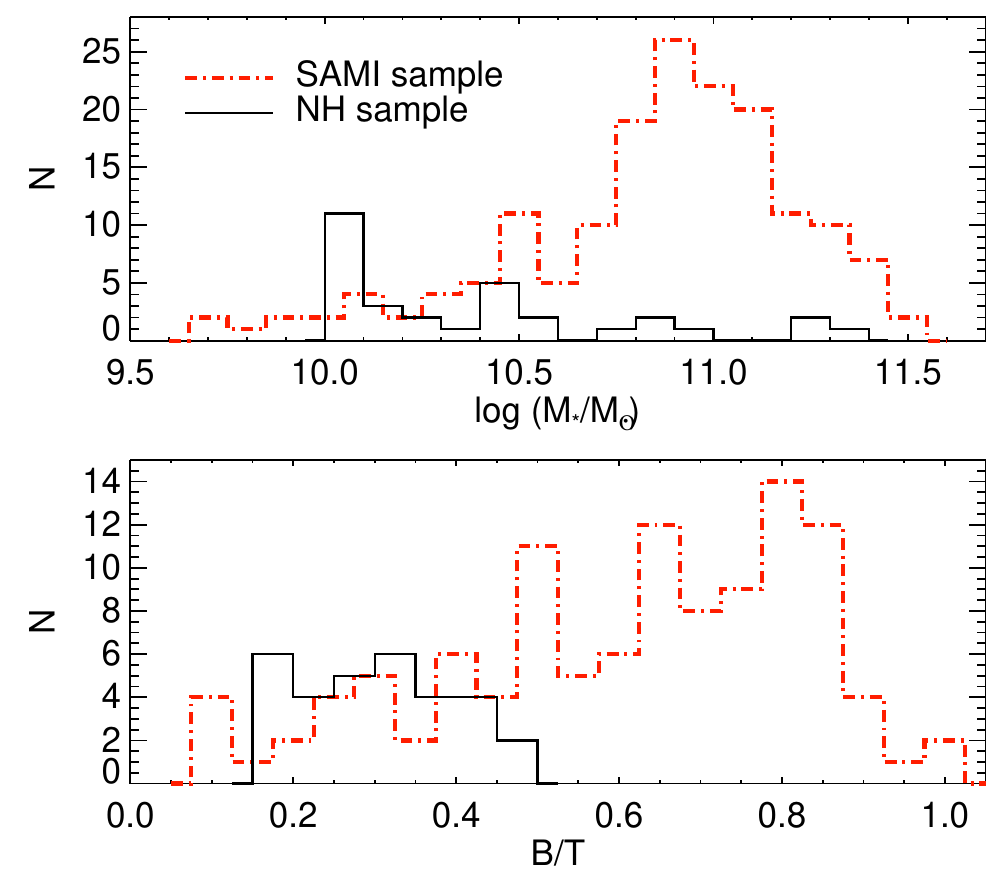}
\caption{The stellar mass and $B/T$ distributions of the 161 SAMI and 31 NH galaxies. While the majority of the SAMI sample are massive and bulge-dominated, the sample also includes galaxies with low $B/T$ values, indicating substantial disc components. The NH sample consists of massive disc galaxies.}
\label{sam_both}
\end{figure}

Santucci et al.\ (2022) constructed orbit-superposition Schwarzschild models (Schwarzschild 1979) for 161 passive SAMI galaxies, selected based on the spectroscopic classification presented in Owers et al. (2019), with stellar masses $\log (M_*/M_\odot) \ge 9.5$, which we adopt in this study. {Although the sample comprises passively evolving systems and tends to be bulge-dominated, a substantial subset still exhibits disc components, spanning a wide range of bulge-to-total light fractions ($0.1 < B/T < 1$), as shown in Figure~\ref{sam_both}.}

On the other hand, galaxies in the NH simulation are predominantly disc-dominated and reside in relatively sparse environments. To focus our analysis on reliably measuring disc structures in the NH simulation, we restrict the sample to 31 massive ($\log (M_*/M_\odot) \ge 10$), disc-dominated galaxies with $B/T < 0.5$. {For comparison, the SAMI sample includes 28 galaxies with $B/T < 0.5$, although SAMI galaxies generally exhibit higher bulge fractions.}

\section{Sampling thin and thick disc component} \label{sec:sampling}
In this section, we describe various methods for sampling the thin and thick disc components and for measuring their kinematics. For SAMI galaxies, we apply a sampling criterion based on the circularity parameter derived from Schwarzschild modelling. For NH simulated galaxies, we explore two sampling schemes: one based on the circularity parameter and the other on stellar particle age.

\subsection{Circularity parameter}
\subsubsection{SAMI: Schwarzschild modelling}
We adopt the orbit-superposition Schwarzschild models constructed by Santucci et al.\ (2022) for 161 passive SAMI galaxies, following the procedures outlined in van den Bosch et al.\ (2008a) and Zhu et al.\ (2018b). Here, we briefly outline the three primary steps involved in Schwarzschild modelling: constructing a model of the gravitational potential; generating a library of orbits within this potential; and determining a combination of orbits that best reproduces the observed kinematics and luminosity distribution. 

The gravitational potential model is constructed using a triaxial stellar component derived from Multi-Gaussian Expansion (MGE; Emsellem et al.\ 1994) luminosity densities, based on $r$-band imaging from the Sloan Digital Sky Survey (SDSS; York et~al.\ 2000) and the VLT Survey Telescope (VST) ATLAS Survey (Shanks et~al.\ 2013). The 3D MGE luminosity model accounts for the galaxy's orientation, defined by three viewing angles, allowing for precise modelling of triaxial shapes. A stellar mass density model is then obtained by applying an estimated mass-to-light ratio ($M/L$). A spherical Navarro-Frenk-White (NFW) dark matter halo is also incorporated, with its mass ($M_{200}$) remaining adjustable to facilitate the construction of orbital model libraries. Each orbit library is sampled across three integrals of motion: energy ($E$), the second integral ($I_2$), and the third integral ($I_3$), using a 21 $\times$ 10 $\times$ 7 point grid. The orbit libraries are then used to reproduce the observed stellar system by adjusting orbit weights. The best-fit model is selected by minimising discrepancies between the modelled and observed luminosity distribution and 2D kinematic maps from the SAMI data. For a more detailed description of the Schwarzschild modelling applied to the SAMI sample, see Santucci et al.\ (2022, 2023).

The orbital circularity parameter ($\lambda_{\rm z}$) characterises the kinematics of orbital families and is commonly used to assess whether stellar orbits are rotationally supported (Zhu et al.\ 2018; Du et al.\ 2019; Park et al.\ 2021; Santucci et al.\ 2022, 2023, 2024; Jang et al.\ 2023; Yi et al.\ 2024). For the SAMI galaxies, the circularity parameter is calculated as follows:
\begin{equation} \lambda_{\rm z, SAMI} = \frac{\overline{L_{\rm z}}}{r \times \overline{V_{\rm c}}} ~, \end{equation}
where $\overline{L_{\rm z}}$ is the time-averaged $z$-component of an orbit's angular momentum, $r$ is the time-averaged orbital radius, and {$\overline{V_{\rm c}}$ is a reference circular-speed term used for normalisation. We therefore interpret $\lambda_{\rm z,SAMI}$ as a proxy for rotational support rather than the circularity of a literal circular orbit in a non-axisymmetric potential (Santucci et al. 2022, Section 4.5).} We adopt the circularity criteria suggested by Du et al.\ (2019) to categorise orbital components into hot ($\lambda_{\rm z,SAMI}< 0.5$), warm ($0.5<\lambda_{\rm z,SAMI}<0.8$), and cold ($\lambda_{\rm z,SAMI}>0.8$) components, which serve as proxies for the bulge, thick disc, and thin disc, respectively. 

\begin{figure}
\centering
\includegraphics[width=\columnwidth]{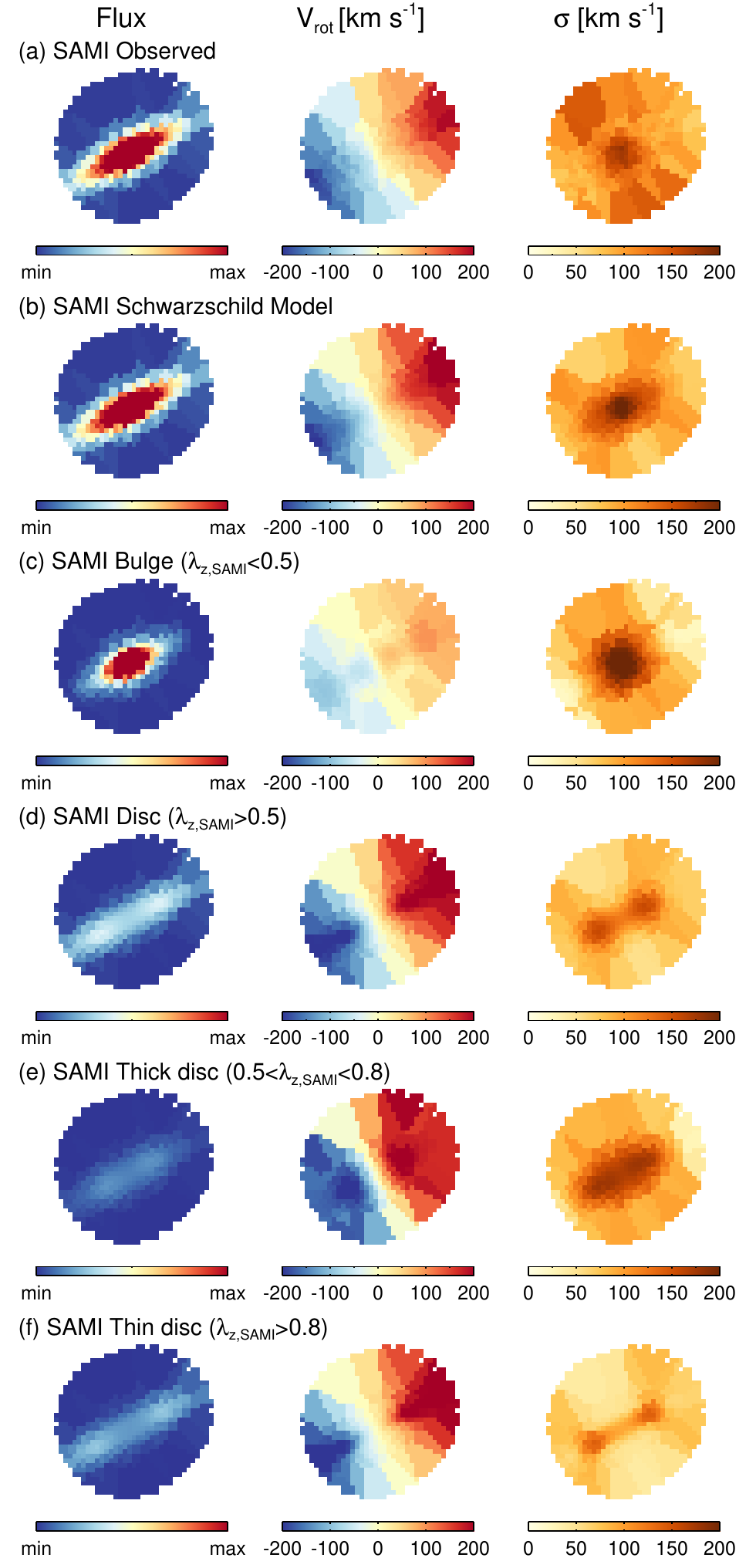}
\caption{Example flux, stellar velocity, and velocity dispersion maps for (a) SAMI observation (ID 278802; $\log (M_*/M_\odot) =10.85$, $B/T=0.33$), (b) the Schwarzschild model reconstruction, and components sampled based on the circularity parameter: (c) bulge, (d) disc, (e) thick disc, and (f) thin disc. }
\label{map_sami}
\end{figure}

Figure~\ref{map_sami} presents example kinematic maps from the SAMI observations alongside those reconstructed by Schwarzschild modelling. We also show the reconstructed kinematic maps for the bulge, disc, thick disc, and thin disc components, sampled based on the circularity parameter.

\subsubsection{NH: the binding energy of a particle} \label{nh_cir}
The circularity parameter for each star particle in the NH galaxies is calculated using the particle binding energy, following Abadi et al.\ (2003):
\begin{equation} \lambda_{\rm z, NH} = \frac{J_{\rm z}}{J_{\rm c}} ~, \end{equation}
where $J_{\rm z}$ is the azimuthal angular momentum of a star particle, and $J_{\rm c}$ is the maximum angular momentum that a particle can attain at a given binding energy $E$. {For consistency in the normalisation used to compute $J_{\rm c}(E)$, we assume a spherically symmetric potential when calculating the binding energy.} We also apply the same sampling criteria for orbital components as used in the SAMI analysis: hot (bulge; $\lambda_{\rm z,NH}< 0.5$), warm (thick disc; $0.5<\lambda_{\rm z,NH}<0.8$), and cold (thin disc; $\lambda_{\rm z,NH}>0.8$).

\begin{figure}
\centering
\includegraphics[width=\columnwidth]{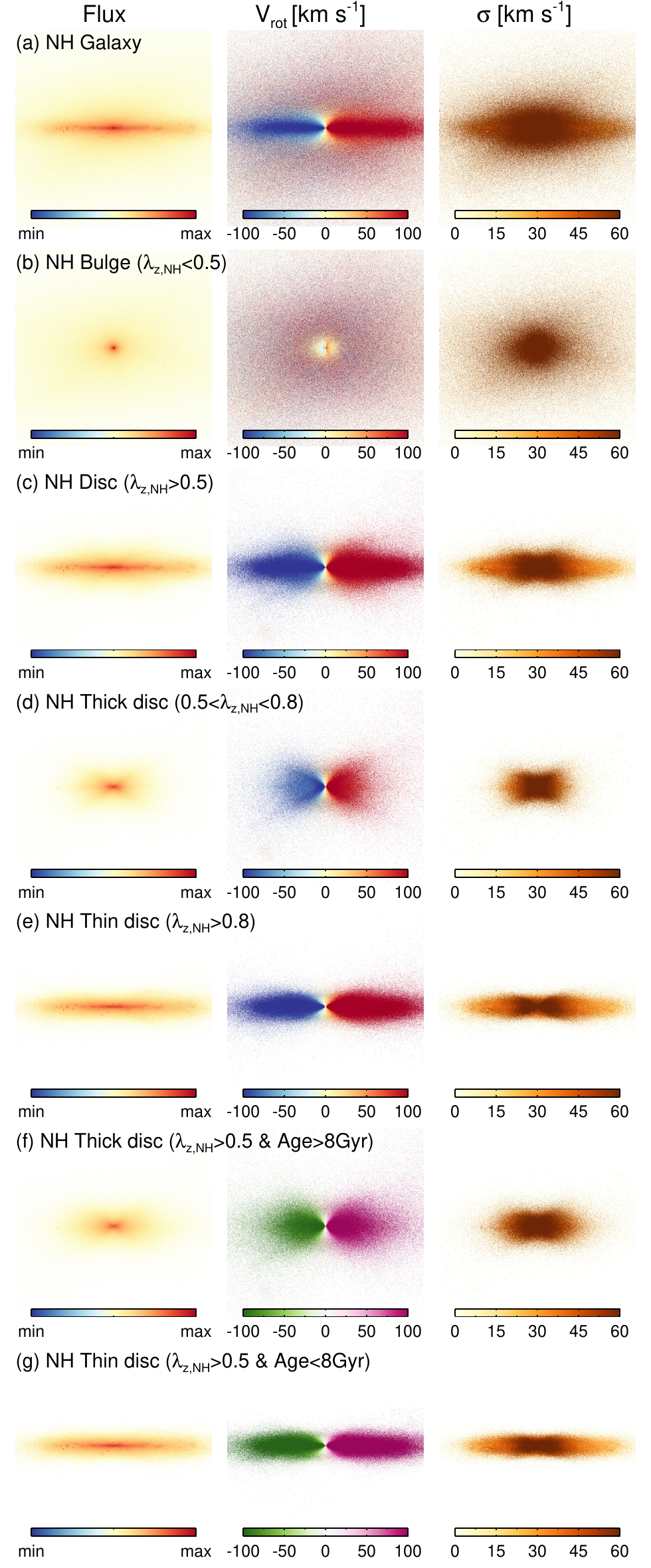}
\caption{Example flux, stellar velocity, and velocity dispersion maps for (a) the NH simulation galaxy 00001 ($\log (M_*/M_\odot) =10.75$, $B/T=0.20$) and its subcomponents, sampled based on the circularity parameter: (b) bulge, (c) disc, (d) thick disc, and (e) thin disc. Panels (f) and (g) show the thick and thin disc kinematics, respectively, based on age-based sampling.}
\label{sam_nh}
\end{figure}

In Figure~\ref{sam_nh}(a)--(e), we present example rotation velocity and velocity dispersion maps for an NH galaxy with a stellar mass of $\log (M_*/M_\odot) =10.75$, along with its orbital components, sampled based on the circularity parameter. This example illustrates the clear kinematic distinction between bulge and disc structures based on orbital dynamics.

\subsection{NH: Age} \label{nh_age}
In addition to the sampling based solely on orbital circularity, we also implement a hybrid method in the NH simulation that incorporates stellar particle age to sample thin and thick disc components. This approach is motivated by evidence that chemically defined thick discs are typically composed of older stars with hotter kinematics and larger scale heights, whereas thin discs host younger, dynamically colder stars (e.g., Reid \& Majewski 1993; Yoachim \& Dalcanton 2006; Bird 2013; Haywood et al.\ 2013; Bensby et al.\ 2014; Hayden et al.\ 2017; Yi et al.\ 2024). Accordingly, stellar age provides a useful parameter for distinguishing between thin and thick disc components, particularly in simulations where formation times are explicitly available.

However, since stellar age alone does not reliably distinguish dynamically hot bulge stars from thick disc stars, we first exclude bulge particles using a circularity threshold of $\lambda_{\rm z, NH} < 0.5$. This also enables a more direct comparison with the thin and thick disc components defined by circularity (Figure~\ref{sam_nh}(d)–(e)), while avoiding contamination from the bulge. We then classify the remaining particles ($\lambda_{\rm z, NH} \ge 0.5$) into thick and thin disc components based on stellar age: particles older than 8 Gyr are assigned to the thick disc, while those younger than 8 Gyr are considered part of the thin disc. This age-based division is supported by the results of Hayden et al.\ (2017), which indicate that the majority of chemically defined thick disc stars in the Milky Way are older than 8 Gyr. Figure~\ref{sam_nh}(f)–(g) presents the resulting kinematic maps of the thick and thin discs obtained through this age-based sampling, illustrating that this approach produces physically distinct components consistent with expectations from galaxy structure.

\section{result}
\subsection{disc \msig\ relation} \label{dmsig}
\begin{figure}
\centering
\includegraphics[width=\columnwidth]{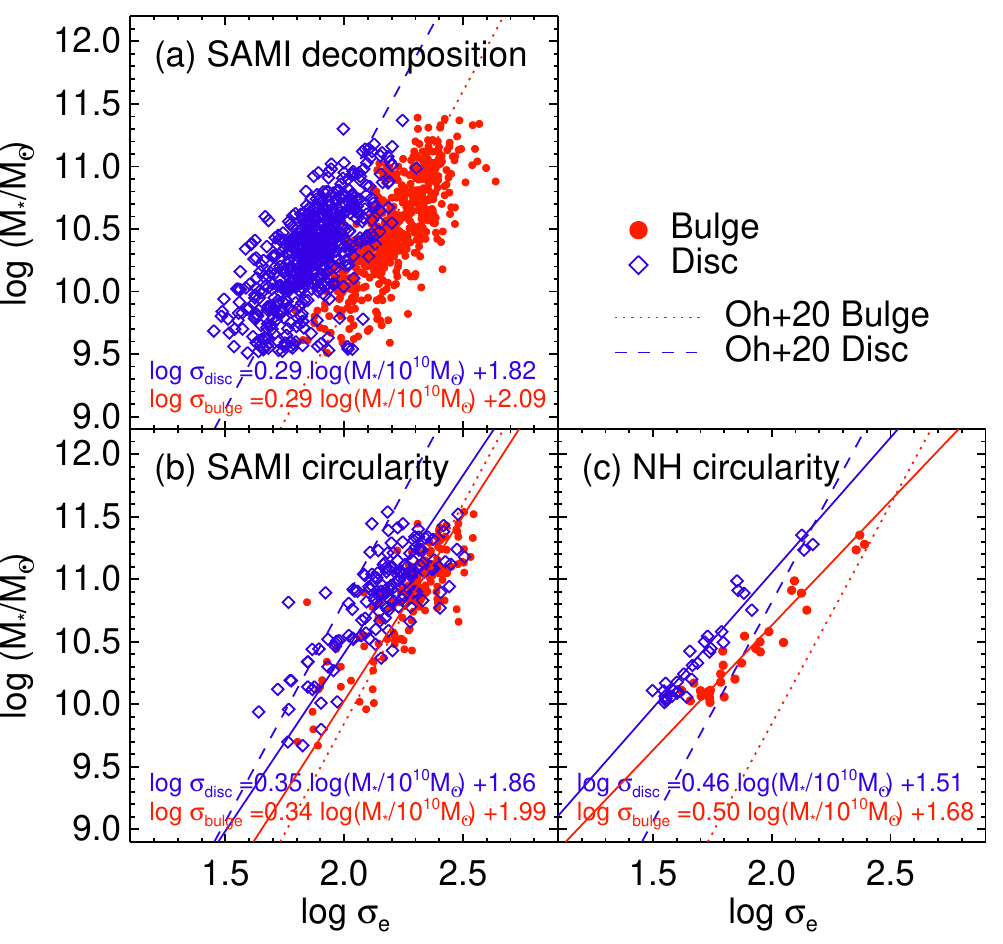}
\caption{Bulge and disc \msig\ relations based on: (a) the SAMI spectroscopic decomposition method from Oh et al.\ (2020); (b) the SAMI circularity parameter derived from Schwarzschild modelling; and (c) the circularity parameter from the New Horizon simulation. {In all cases, the total stellar mass is shown, and \esig\ is measured separately for each component.} In panel (a), bulges and discs are spectroscopically separated, with the dotted and dashed lines indicating the least-squares fits to the bulge and disc components, respectively. Panels (b) and (c) use orbital circularity to classify components, where bulges are defined as $\lambda_{\rm z} < 0.5$ and discs as $\lambda_{\rm z} > 0.5$. {The solid lines indicate the best-fit \msig\ relations for each component, with the corresponding fitting equations shown at the bottom of each panel.} The disc \msig\ trends are consistently offset from, but nearly parallel to, the bulge relation across all three methods, reflecting underlying structural differences despite variations in slope and normalisation resulting from differences in data and classification criteria.}
\label{msig}
\end{figure}

In this section, we revisit the stellar mass–velocity dispersion (\msig) relation for disc components, as originally presented in Oh et~al.\ (2020). Figure~\ref{msig}(a) reproduces the scaling relation between \emph{total} stellar mass and the flux-weighted mean stellar velocity dispersion measured within 1\re\ for the bulge and disc components, as shown in Figure 8 of Oh et~al.\ (2020). The underlying dataset comprises 826 galaxies from the SAMI survey, spectroscopically decomposed using \textsc{pPXF} (Cappellari \& Emsellem 2004; Cappellari 2017), with photometric weights from Barsanti et~al.\ (2021) and Casura et~al.\ (2022). Notably, the disc component exhibits a well-defined \msig\ relation nearly parallel to that of the bulge component or elliptical galaxies, although the \esig\ for discs at fixed $M_*$ is lower by approximately a factor of two. Furthermore, the \esig\ scatter in the \msig\ relation for the disc component is comparable to that of both the bulge component and early-type galaxies (see Table 1 of Oh et~al.\ 2020). {While beam-smearing can, in principle, affect \esig\ measurements, especially for galaxies with steep central rotation gradients, Oh et~al.\ (2020) show that the \msig\ relation for bulge and disc components remains robust even when galaxies significantly affected by beam-smearing are excluded.}

In Figure~\ref{msig}(b), we present the \msig\ relation for the bulge ($\lambda_{\rm z,SAMI} < 0.5$) and disc ($\lambda_{\rm z,SAMI} > 0.5$) components, as kinematically defined by the Schwarzschild modelling of the 161 SAMI sample. Velocity dispersions are measured as flux-weighted means within 1\re\ from the model velocity dispersion maps, using the same approach as in Figure~\ref{msig}(a). The bulge \msig\ relation agrees well with that obtained from the spectroscopic bulge–disc decomposition. The disc component also exhibits a well-defined \msig\ relation with a nearly parallel slope, though offset by approximately 0.2 dex toward higher \esig\ at fixed $M_*$. We discuss possible origins of this offset in Section~\ref{sec:sch_offset}.

Simulation studies support the existence of a tight disc \msig\ relation. Irodotou \& Thomas (2021) used angular momentum maps to decompose discs and spheroids in EAGLE simulation galaxies (Schaye et al.\ 2015; Crain et al.\ 2015), reporting well-constrained \msig\ relations for both components, with slopes consistent with those observed by Oh et al.\ (2020).

We further examine the bulge and disc \msig\ relations using galaxies from the NH simulation. Components are identified using the orbital circularity parameter, applying the same threshold as in the SAMI Schwarzschild analysis to identify hot (i.e., bulge; $\lambda_{\rm z, NH} < 0.5$) and cold (i.e., disc; $\lambda_{\rm z, NH} > 0.5$) orbital families. {The velocity dispersion is defined as \esig\ = $\sqrt{(\sigma_r^2 + \sigma_t^2 + \sigma_z^2)/3}$, computed from stellar particles within 1\re\ (i.e., mass-based, not flux-weighted). Constructing a strictly flux-weighted analogue would require additional assumptions (e.g., band-dependent mass-to-light ratio and dust attenuation), so we retain the mass-based definition for the simulation.} As shown in Figure~\ref{msig}(c), NH galaxies show clear \msig\ relations for both bulge and disc components, though with shallower slopes and lower \esig\ at fixed $M_*$. The origin of this systematic offset is discussed in Section~\ref{sec:nh_offset}.

Both observational and theoretical studies confirm that the disc component exhibits an \msig\ relation comparable to that of the bulge component (or elliptical galaxies) in both slope and tightness (see Oh et~al.\ 2020). While disc components are primarily supported by rotation, leading most kinematic studies to focus on the scaling between mass and rotation velocity (i.e., the Tully–Fisher relation), the presence of a reasonably tight \msig\ relation for discs highlights the need to examine their velocity dispersion and understand the physical origin of this correlation. In the following section, we further decompose the disc components into thin and thick components, which have been recognised through decades of observational studies as possessing distinct characteristics, in order to investigate their individual scaling relations with velocity dispersion.

\subsection{Thin and thick disc scaling relation}
As described in Section~\ref{sec:sampling}, we identify the thin and thick disc components for the SAMI sample based on the circularity parameter obtained from Schwarzschild modelling. The velocity dispersions of the thin and thick discs ($\sigma_{\rm tn,\lambda_z,SAMI}$ and $\sigma_{\rm tk,\lambda_z,SAMI}$) are measured as flux-weighted means within 1\re, using the corresponding velocity dispersion maps. {For the NH galaxies, we compute the velocity dispersion within 1\re\ as \esig\ = $\sqrt{(\sigma_r^2 + \sigma_t^2 + \sigma_z^2)/3}$ in cylindrical coordinates (mass-based). We consider two sampling criteria for selecting thin and thick components}: the circularity parameter based on the particle’s binding energy (Section~\ref{nh_cir}; $\sigma_{\rm tn,\lambda_z,NH}$ and $\sigma_{\rm tk,\lambda_z,NH}$), and stellar age with a division at 8 Gyr (Section~\ref{nh_age}; $\sigma_{\rm tn,age,NH}$ and $\sigma_{\rm tk,age,NH}$).

\begin{figure}
\centering
\includegraphics[width=\columnwidth]{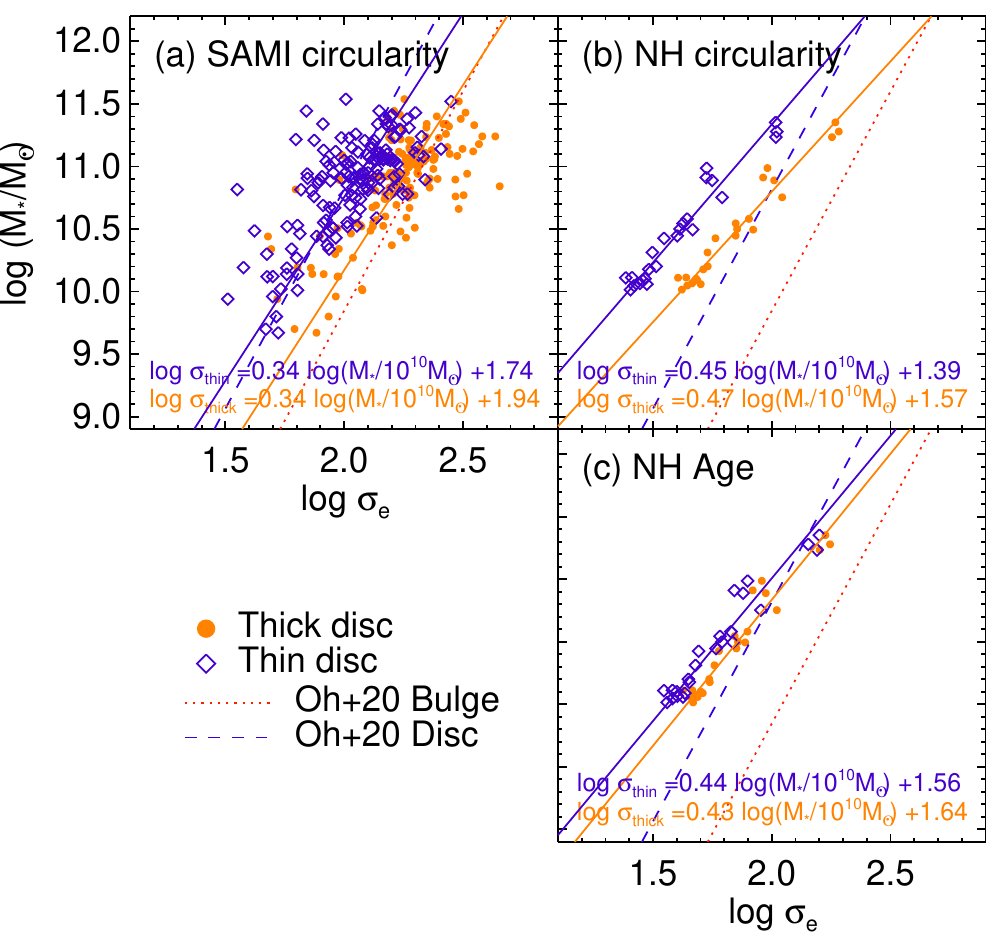}
\caption{Thin (diamonds) and thick (circles) disc \msig\ relations based on: (a) the circularity parameter derived from Schwarzschild modelling of the SAMI sample; (b) the circularity parameter from the New Horizon simulation; and (c) the stellar age of the particles in the New Horizon simulation. {In all cases, the total stellar mass is presented. The solid lines indicate the best-fit \msig\ relations for each component, with the corresponding fitting equations shown at the bottom of each panel.}  The dotted and dashed lines indicate the bulge and disc \msig\ relations from Oh et al.\ (2020), shown for comparison. }
\label{disc_msig}
\end{figure}

In Figure~\ref{disc_msig}, we present the \msig\ relations for the thin and thick disc components in both the SAMI and NH samples, classified according to different sampling criteria. In all cases, the thin and thick discs exhibit reasonably well-defined \msig\ relations, with the thick discs generally showing higher $\sigma$ values at fixed $M_*$ than the thin discs. {We note, however, that the exact offset, slope, and scatter depend on the specific choice of circularity or age thresholds used to separate the cold (thin) and warm (thick) components.} Systematic differences in slope and offset between the SAMI Schwarzschild modelling and the NH simulation, already apparent in the overall disc \msig\ relations (Figure~\ref{msig}), persist after separating the thin and thick disc components and are further discussed in Section~\ref{sec:nh_offset}.

\begin{figure*}
\centering
\includegraphics[width=\textwidth]{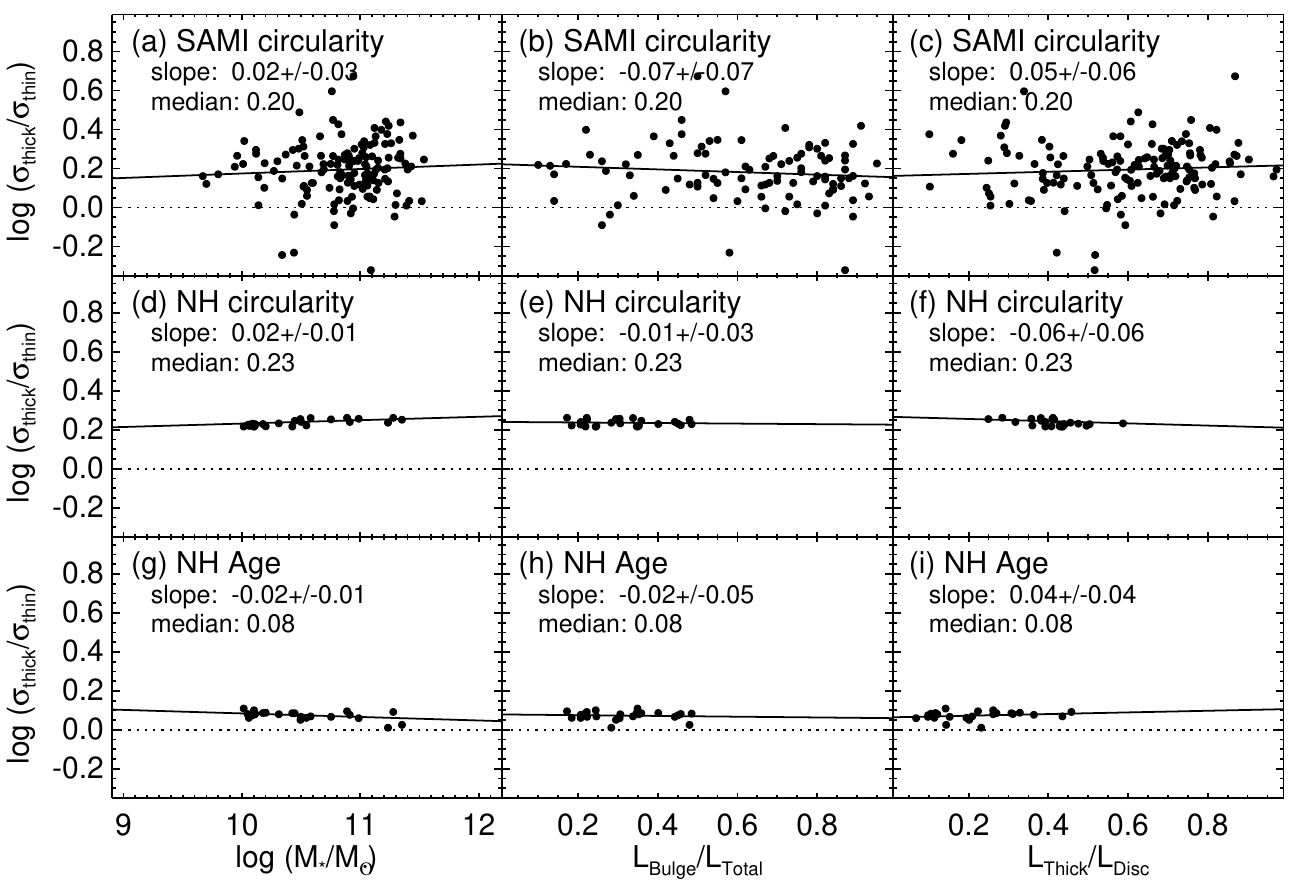}
\caption{Logarithmic ratio of thick-to-thin disc velocity dispersions as a function of stellar mass (left), bulge-to-total luminosity ratio (middle), and thick-to-disc luminosity ratio (right). Panels (a–c) show results based on the circularity parameter from Schwarzschild modelling of the SAMI sample; (d–f) use the circularity parameter from the New Horizon simulation; and (g–i) are based on stellar age from the New Horizon simulation. Each panel indicates the median value and the best-fit linear slope with its 1$\sigma$ uncertainty. Across all panels, the velocity dispersion ratio remains nearly constant, with median values of $\sim$0.20 for SAMI and $\sim$0.23 for NH (circularity-based), and a lower value of $\sim$0.08 for NH (age-based), highlighting a weaker kinematic distinction when age is used as the classification criterion.}
\label{disc_res}
\end{figure*}

To minimise the influence of systematic differences between datasets and sampling methods, and to better highlight the intrinsic kinematic contrast between the thin and thick discs, Figure~\ref{disc_res} presents the logarithmic ratio of their velocity dispersions, $\log(\sigma_{\rm thick}/\sigma_{\rm thin})$, as a function of stellar mass, bulge and thick disc luminosity fractions. The median values are consistent across the SAMI and NH samples when using the circularity-based classification (median $\sim$ 0.20 to 0.23; Figure~\ref{disc_res}a–f). {Since the components are selected by circularity, the exact dispersion contrast depends on the adopted thresholds (see Section~\ref{circularity_selection}). Nonetheless, the agreement between observations and simulations indicates that, under matched selection, both datasets produce a similar mapping between orbital circularity and kinematic temperature.}

The age-based classification in NH yields a smaller median ratio of 0.08 (Figure~\ref{disc_res}g--i), suggesting that the kinematic distinction between the thin and thick components is less pronounced when the separation is defined by stellar age. {This highlights the sensitivity of the inferred kinematic properties to the adopted component classification scheme. {Alinder et al. (2025) provide Milky Way context by showing that inferred thin/thick disc properties can depend on the adopted labelling scheme, consistent with the label dependence we highlight here.} 

In this work, we treat the circularity-based decomposition as a purely kinematic definition and the age-based sampling as an independent, physically motivated alternative. Although the absolute contrast in velocity dispersion between the thin and thick components depends on the chosen classification, both approaches consistently indicate that the thicker component is dynamically hotter and that the disc components follow a well defined stellar mass velocity dispersion scaling.} While slight variations in slope are observed, all panels in Figure~\ref{disc_res} show statistically consistent flat trends within the uncertainties, implying that the ratio of thick to thin disc velocity dispersion remains approximately constant across galaxy mass and type.

\subsection{Sensitivity to circularity-based separation criteria} \label{circularity_selection}
\begin{figure}
\centering
\includegraphics[width=\columnwidth]{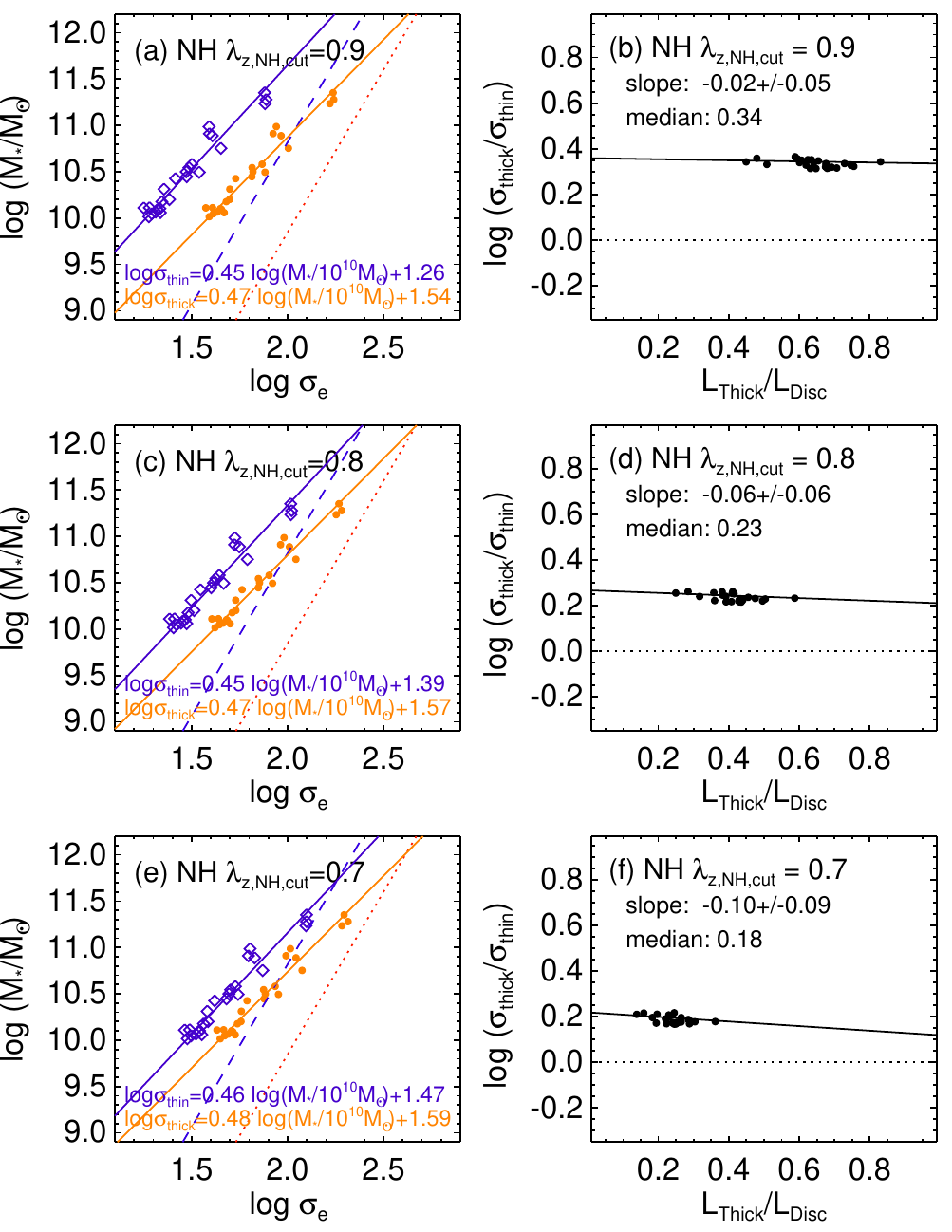}
\caption{{Thin and thick disc \msig\ relations, together with the corresponding $\log(\sigma_{\rm thick}/\sigma_{\rm thin})$ values as a function of thick-to-disc luminosity ratio, for different circularity thresholds. Details are the same as in Figures~\ref{disc_msig} and~\ref{disc_res}. Panels (a) and (b), (c) and (d), and (e) and (f) show the results for $\lambda_{\rm z,NH,cut}=0.7$, 0.8 (fiducial), and 0.9, respectively. In each case, the thick disc is defined by $0.5<\lambda_{\rm z,NH}<\lambda_{\rm z,cut}$, while the thin disc is defined by $\lambda_{\rm z,NH}>\lambda_{\rm z,cut}$. With increasing $\lambda_{\rm z,NH}$, the inferred thick disc fraction becomes larger and the median thick-to-thin dispersion ratio also increases.}}
\label{disc_msig_cir}
\end{figure}

{The separation between thin and thick disc components depends on the adopted circularity threshold, so it is useful to check how strongly our results depend on this choice. Throughout this paper, we use $\lambda_{z,{\rm NH,cut}} = 0.8$ as our fiducial threshold for separating the thin and thick discs. Here, we test the sensitivity of our results by repeating the analysis for $\lambda_{z,{\rm NH,cut}} = 0.7$ and 0.9 (Figure~\ref{disc_msig_cir}).}

We find that the mass dependence of the thin- and thick-disc \msig\ relations is robust to the adopted circularity threshold. In particular, the fitted \msig\ slopes remain nearly unchanged across all choices, indicating that the scaling of disc kinematics with stellar mass is not driven by the specific circularity selection.

{In contrast, the normalization of the kinematic separation does vary with the adopted threshold. As the circularity cut increases, the thick disc component becomes increasingly dominant, while the thin disc component becomes systematically colder. As a result, the median $\log\sigma_{\rm thick}/\sigma_{\rm thin}$ increases from 0.18 to 0.34. This trend is expected, because increasing the circularity threshold makes the thin disc selection more restrictive, leaving only the coldest orbits in the thin disc component, while the velocity dispersion of the thick-disc component remains nearly unchanged. Despite this shift, the dispersion ratio still shows only a weak dependence on stellar mass and thick-disc fraction for all threshold choices. This suggests that while the absolute level of kinematic separation is sensitive to the adopted circularity definition, whereas the overall scaling with galaxy properties remains similar.}

\subsection{Sensitivity to age-based separation criteria} \label{age_selection}
In our main analysis, we adopted an age threshold of 8\,Gyr, motivated by the results of Haywood et al.\ (2013), who showed that the Milky Way’s thin and thick disc populations are cleanly divided at around 8\,Gyr. However, other studies such as Hayden et al.\ (2017) suggest that the separation between the two components is less distinct in age space. Moreover, it is uncertain whether the Milky Way is representative of disc galaxies in general, particularly when interpreting simulation data (see Figure~A6 of Yi et al.\ 2024).

\begin{figure}
\centering
\includegraphics[width=\columnwidth]{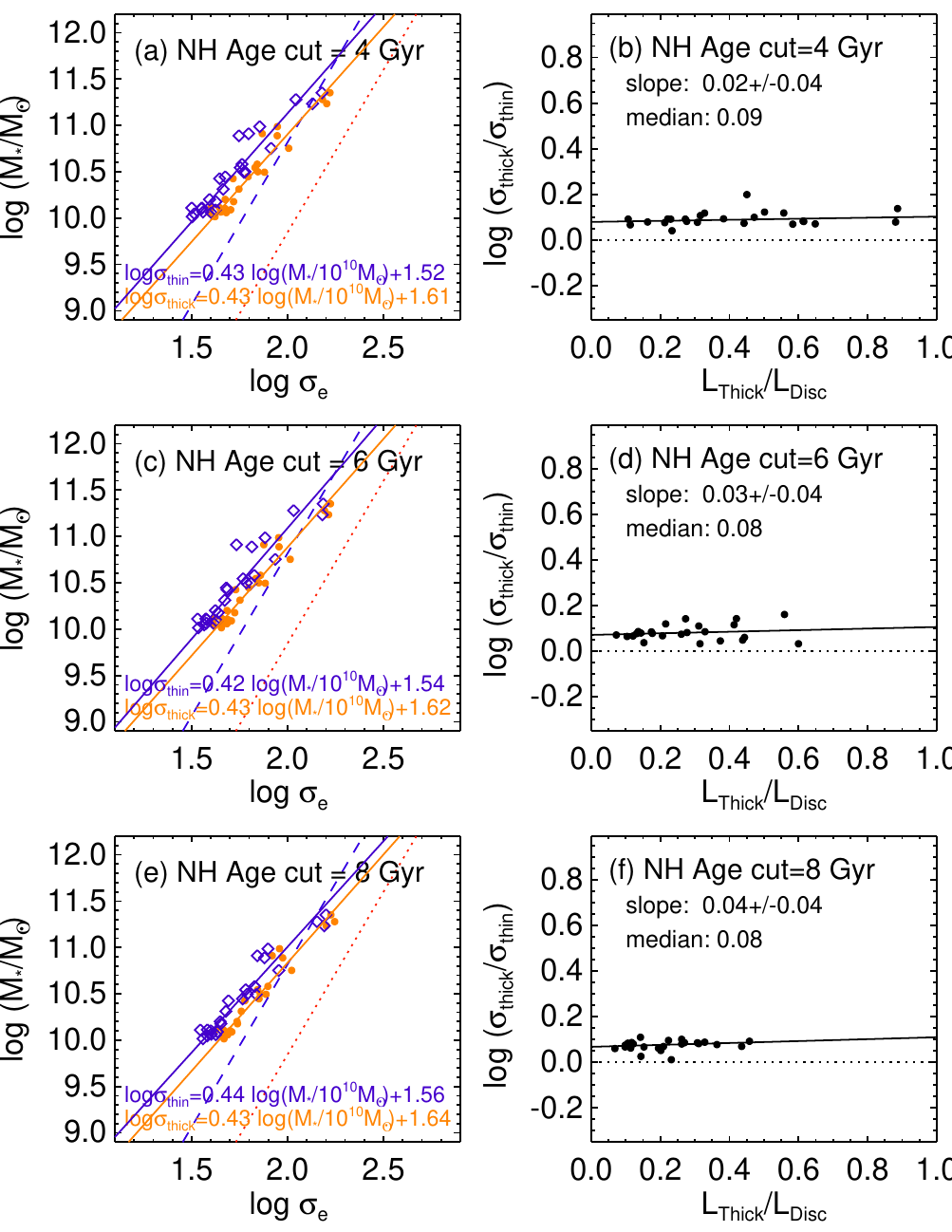}
\caption{Thin and thick disc \msig\ relations and corresponding $\log(\sigma_{\rm thick}/\sigma_{\rm thin})$ values as a function of thick-to-disc luminosity ratio, for different stellar age thresholds. Details are the same as in Figures~\ref{disc_msig} and~\ref{disc_res}. Panels (a) and (b) use a threshold of 6\,Gyr, (c) and (d) adopt 8\,Gyr (fiducial), and (e) and (f) use 10\,Gyr to separate the thin and thick disc components. Across all cases, the trends and median values remain nearly unchanged, indicating that the kinematic contrast between the thin and thick discs is robust against reasonable variations in the age-based classification scheme.}
\label{disc_msig_age}
\end{figure}

To evaluate the robustness of our kinematic results to this age classification criterion, we tested two additional age thresholds: 4 and 6\,Gyr, in addition to the fiducial 8\,Gyr. {We note that the NH simulation is analyzed at $z=0.17$, so the particle ages used here correspond to that epoch.
For reference, age thresholds of 4, 6, and 8\,Gyr at $z=0.17$ would correspond to approximately 6.1, 8.1, and 10.1,Gyr, respectively, if evolved to $z=0$.} Figure~\ref{disc_msig_age} shows the resulting logarithmic velocity dispersion ratio, $\log(\sigma_{\rm thick}/\sigma_{\rm thin})$, for each case. Across all thresholds, the trends and median values remain consistent, indicating that the overall conclusions are insensitive to the precise choice of the age boundary.

{Across the range of plausible age selections, the inferred dispersion contrast between the two age-defined disc components remains moderate. This reflects the fact that a single global age cut does not necessarily provide a uniform separation of dynamically distinct disc material across the full stellar-mass range. We discuss this connection between age and kinematics in more detail in Section 5.3.}

\section{Discussion}
\subsection{Systematic offset in disc \esig\ between Schwarzschild modelling and spectroscopic bulge--disc decomposition} \label{sec:sch_offset}
When sampling the disc component based on the circularity of stellar orbits from the Schwarzschild modeling, we define the disc as the combination of the cold ($\lambda_{\rm z,SAMI} > 0.8$) and warm ($0.5<\lambda_{\rm z,SAMI} < 0.8$) components. However, the disc velocity dispersion derived from the Schwarzschild modelling is found to be significantly larger than that obtained from the spectroscopic bulge--disc decomposition (see Figure~\ref{msig}a and b), which uses the photometric $B/T$ as a prior constraint. This discrepancy likely stems from the fundamental difference in how discs and bulges are identified in each method. {As a result, the mapping between the Schwarzschild cold plus warm orbit families and the spectroscopically decomposed disc component is not one-to-one, especially in bulge-dominated systems.} Specifically, the photometric decomposition uses light profiles to define the disc and bulge, whereas the Schwarzschild modeling relies on orbital classifications, separating stellar components into cold, warm, and hot populations based on their kinematics.

Zhu et al.\ (2018) conducted a detailed analysis of 250 galaxies from the CALIFA survey using Schwarzschild modeling. They found that the behavior of the kinematically warm component varies systematically with galaxy morphology when compared to the photometric $B/T$ ratio. Specifically, for disc-dominated galaxies, the combined contribution of the cold and warm components closely matches the photometric disc fraction. In contrast, for early-type galaxies, the cold component alone is more consistent with the photometric disc fraction, while including the warm component leads to a significantly higher inferred disc fraction than that derived from photometry.

Given that majority of the galaxies in our Schwarzschild modeling sample are bulge-dominated systems (Section~\ref{sec:sam}), and that the disc is defined in our analysis as the combination of the cold and warm components, {the Schwarzschild-based disc velocity dispersion can be larger than that obtained from spectroscopic decomposition.} This difference can arise because the warm component, which can have a velocity dispersion even comparable to that of the bulge, is included as part of the disc in the Schwarzschild-based approach. In contrast, the photometric disc component used as a prior in the spectroscopic decomposition likely excludes such dynamically warm stars, resulting in a systematically lower measured disc velocity dispersion. We also note that photometric decompositions that enforce a pure exponential disc can assign bars or other central disc related structures to the bulge, which may further shift the photometric disc toward dynamically colder stars.

This interpretation is further supported by the \msig\ relation for the thin disc (i.e. kinematically cold) component (Figure~\ref{disc_msig}a), which shows relatively good agreement with the disc \msig\ relation derived from the spectroscopic decomposition (Figure~\ref{msig}a). {This consistency suggests that the kinematically cold component corresponds more closely to the photometric disc, while the warm component is not always captured by purely photometric definitions of the disc, in line with the results of Zhu et al.\ (2018).}

{As a check, we compared the Schwarzschild-based disc ($\log\,$\esig) with the value expected from the disc (\msig) relation derived from the spectroscopic decomposition of Oh et al.\ (2020) at the same stellar mass (Figure~\ref{msig}a; log\,\esig = 0.29 log $(M_*/10^{10}M_{\odot})+1.82$). By construction, an offset of zero would indicate consistency with the spectroscopic-decomposition disc relation, while positive values indicate higher Schwarzschild-based disc dispersions. The median offset is larger for bulge-dominated galaxies ($B/T>0.5$, 0.11 dex) than for disc-dominated galaxies ($B/T<0.5$, 0.06 dex). This trend is in the expected direction, suggesting that bulge dominance contributes to the discrepancy. However, the offset remains non-zero even for disc-dominated galaxies, indicating that morphology alone does not fully explain the difference. The offset therefore likely reflects a combination of morphology-dependent effects and the different definitions of the disc component in the Schwarzschild-based and spectroscopic decomposition approaches.}

\subsection{Systematic offset in the \msig\ relation between observation and simulation} \label{sec:nh_offset}
On the other hand, NH galaxies exhibit lower \esig\ values for both bulge and disc components at fixed stellar mass, particularly in the low-mass regime, resulting in a shallower slope in the \msig\ relation compared to that derived from the SAMI spectroscopic decomposition (Figure~\ref{msig}a and c). {The comparison between SAMI and NH is not a strictly morphology-matched test of individual galaxy populations, because the Oh et al.\ sample shown in Figure~\ref{msig}(a) includes galaxies spanning a wide range of morphological types with $M_* > 10^{9.5},M_\odot$, whereas the NH sample (Figure~\ref{msig}c) is limited to disc-dominated galaxies with $M_* > 10^{10},M_\odot$. Nevertheless, the difference in sample selection alone is unlikely to explain the discrepancy in velocity dispersion between the simulation and observation. Oh et al.\ (2020) demonstrated that the \msig\ relations of the decomposed bulge and disc components remain broadly consistent across different morphological types. Consistent with this, we do not find any significant morphological dependence in Figures~4,~5, and 6 for the SAMI sample. For example, after restricting the SAMI sample in Figure~\ref{msig} to match the NH mass and morphology selection ($M_* > 10^{10}M_\odot$ and $B/T < 0.5$), the median offset of the SAMI disc $\log\,$\esig\ values relative to the NH disc \msig\ relation at the same stellar mass remains approximately 0.25 dex.}


We note that although the NH simulations correspond to a slightly higher redshift ($z \sim 0.17$) than the SAMI sample ($z < 0.095$), this modest offset is unlikely to introduce significant biases in the derived scaling relations. Moreover, several studies investigating stellar velocity dispersions in simulations have reported that simulated galaxies exhibit lower velocity dispersions at a given stellar mass compared to observed galaxies (van de Sande et al.\ 2019; Lu et al.\ 2020; Ferrero et al.\ 2021; Irodotou \& Thomas 2021; Sohn et al.\ 2024). {Although a direct comparison is challenging because velocity dispersion estimates depend on aperture definitions, weighting schemes, and observational effects, the existing studies point to a persistent tension between simulations and observations.}

van de Sande et al.\ (2019) compared the \msig\ relations using mock observations from several cosmological simulations—\textsc{eagle} (Crain et al.\ 2015; Schaye et al.\ 2015), \textsc{hydrangea} (Bah\'e et al.\ 2017), \textsc{horizon-agn} (Dubois et al.\ 2014), and the \textsc{magneticum pathfinder}$^4$ (Teklu et al.\ 2015)—with IFS observations from ATLAS$^{3D}$ (Cappellari et al.\ 2011), CALIFA (S\'anchez et al.\ 2012), SAMI, and MASSIVE (Ma et al.\ 2014). They found that the simulations systematically underestimate the velocity dispersion at a fixed stellar mass by 0.15–0.23 dex, suggesting that the discrepancy in velocity dispersion does not originate from differences in methodology, since mock observations were applied consistently to the simulation data. They argued that this discrepancy is unlikely to be caused by observational seeing (i.e.\ beam smearing) effects. Nevertheless, this possibility should still be rigorously tested to eliminate any remaining questions about the potential artificial inflation of the observed \esig\ due to seeing and rotational contributions (e.g.\ Varidel et al.\ 2019; Oh et al.\ 2022), particularly given the good agreement observed in the baryonic Tully–Fisher relation of NH galaxies (Figure 20 of Dubois et al. 2021). 

Even aside from seeing effects, the clean separation of rotational and random motions, which is readily achieved in simulations, is difficult to reproduce in observations, which may lead to some level of enhancement in the measured velocity dispersions. Differences in weighting schemes may influence the measured velocity dispersions and the slopes of the scaling relations. In general, flux-weighted observational estimates tend to be biased toward younger and dynamically colder stellar components, which would result in lower velocity dispersions compared to mass-weighted values from simulations. However, since our simulations already yield lower dispersions than the observations, differences in weighting alone are unlikely to be the primary driver of the observed offset. Another factor worth investigating is the effect of spectral resolution, which remains moderate in current IFS datasets. The upcoming Hector survey (Bryant et al.\ 2024; Oh et al.\ 2025), with its higher spectral resolution, will provide an opportunity to assess the influence of spectral resolution on measured stellar kinematics.

Ferrero et al.\ (2021) reported that elliptical galaxies in the TNG and EAGLE simulations exhibit systematically lower velocity dispersions at fixed stellar mass compared to observations, resulting in an offset in the Faber--Jackson relation. They attributed this discrepancy to the overly large sizes of low-mass ellipticals in simulations and the formation of massive ellipticals in halos less massive than expected, both of which contribute to the offset in the \msig\ relation across the entire mass range.

It is possible that current simulations do not fully capture the turbulent motions of gas and stars driven by processes such as stellar feedback and AGN feedback. This may result in simulated galaxies being more dynamically settled and kinematically colder than observed, potentially leading to lower velocity dispersions.

{In addition to the offset in normalisation, we also note that NH shows markedly smaller scatter than SAMI, which may reflect both observational and simulation-dependent effects. On the observational side, measurement uncertainties, projection effects, and decomposition systematics can broaden the SAMI relations. On the simulation side, the reduced scatter in NH may partly reflect subgrid prescriptions calibrated primarily to mean galaxy relations and the relatively quiet environment selected for the NH zoom-in, which may under-sample rare heating events and reduce galaxy-to-galaxy diversity.}

Although we find systematic differences in velocity dispersions between observations and simulations, making it difficult to directly compare the slope or intercept of the kinematic scaling relations, the internal comparison between thin and thick disc velocity dispersions presented in this study remains meaningful. This is because many of these systematic effects are, to first order, common to subcomponents measured within the same dataset and with the same methodology.

\subsection{Thin and thick disc definition}
In this section, we discuss the classification criteria based on kinematics and stellar age, and the key physical properties that differentiate the two disc components. Overall, we find a thick-to-thin disc velocity dispersion ratio of $\sigma_{\rm thick}/\sigma_{\rm thin}\sim1.6$ in both the SAMI and NH samples when discs are classified by orbital circularity (Figure~\ref{disc_res}). {Because the separation is defined kinematically, this ratio is not independent of the adopted circularity thresholds (Figure~\ref{disc_msig_cir}). Even so, the agreement between the SAMI and NH samples indicates that the two datasets are consistent when the same circularity based definition is applied.} Given that vertically thicker structures are more kinematically heated and thus have higher mean velocity dispersions, orbital circularity offers a practical proxy for separating dynamically colder and hotter disc material that aligns with the traditional thick and thin disc picture. Age-based classifications also yield a robust \msig\ relation, with a hotter thick disc and a colder thin disc, confirming a link between kinematics and the stellar-population ages of the two components (see also Haywood et al.\ 2013; McCluskey et al.\ 2025). However, the kinematic separation is modest ($\sigma_{\rm thick}/\sigma_{\rm thin}\sim1.2$) relative to the circularity-defined components, indicating that the two approaches are related but not equivalent.

\begin{figure}
\centering
\includegraphics[width=\columnwidth]{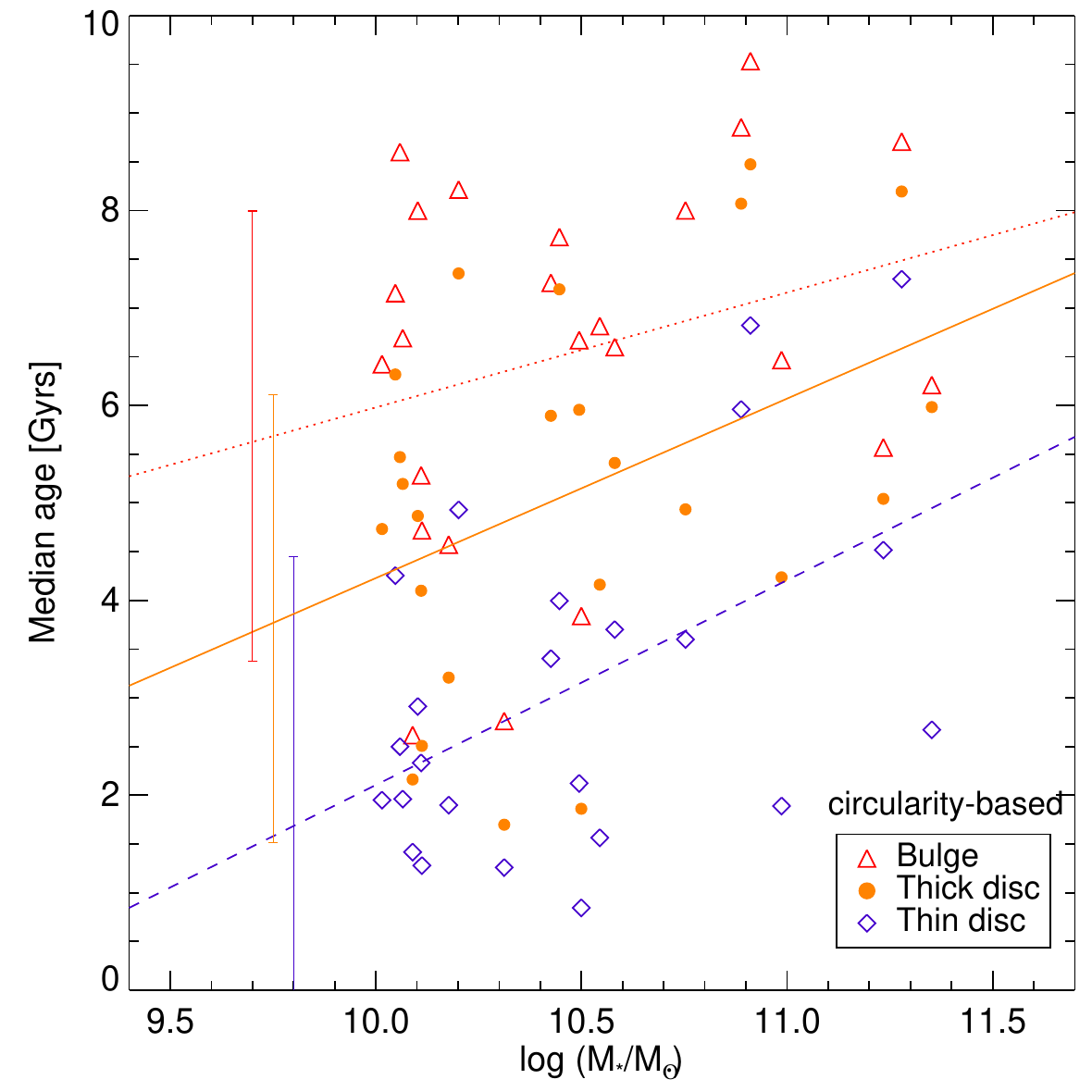}
\caption{Median stellar age versus stellar mass in the New Horizon simulation for bulge (triangles), thick disc (circles), and thin disc (diamonds) components, with discs defined by orbital circularity. The dotted, solid, and dashed lines show least-squares fits to the bulge, thick-disc, and thin-disc sequences, respectively. Error bars indicate the typical 1~$\sigma$ distribution in age for each component. {Note that the simulation is analyzed at $z=0.17$, so the stellar ages shown here correspond to that epoch. For reference, they would be approximately 2.1\,Gyr older if passively evolved to $z=0$.}}
\label{disc_age}
\end{figure}

We further investigate circularity- and age-based classifications by examining the median ages of the circularity-defined thick and thin components (Figure~\ref{disc_age}). {The circularity-based selection yields an age ordering in the median across the mass range, with bulges oldest and thick discs older than thin discs.} At fixed stellar mass, the age distributions are broad and substantially overlapping, but the thick discs are nevertheless systematically older than the thin discs across the sample. If circularity is taken as the reference definition, a single age threshold (for example, 6 or 8 Gyr) cannot reproduce a comparable thick–thin disc split over the full mass range, because the median ages of all components increase with mass (Figure~\ref{disc_age}). A global age cut therefore yields different memberships between the two disc components, particularly thin-disc material in high-mass systems and thick-disc material in low-mass systems. This explains why age-based selections yield only a modest kinematic separation and are not equivalent to circularity-based classifications.

The strong [$\alpha$/Fe]–age correlation observed in the Milky Way (Haywood et al.\ 2013) motivates extending our discussion to identifying the two disc components via chemical composition. Recently, Yi et al.\ (2024) analysed the NH and NewHorizon2 (NH2) simulations, which share the same initial conditions; NH2 additionally tracks detailed chemical abundances. They kinematically and chemically sampled thick discs and showed that a single [$\alpha$/Fe] threshold both contaminates the “chemically thick” sample with thin-disc stars and misses many kinematically thick stars. They also reproduced a positive [$\alpha$/Fe]–age correlation in NH2. Our finding that age- and circularity-based classifications yield different outcomes is therefore consistent with their results.

Assuming that the thin and thick discs are commonly distinguishable substructures in spiral galaxies (analogous to bulge versus disc), differences between kinematic and age/chemical classifications raise the question of which parameter best discriminates between them. Although the two approaches often overlap, they probe different aspects of galaxy evolution: age/chemical schemes emphasise formation history and stellar population origin, whereas kinematic schemes reflect the present-day dynamical state by separating stars on hotter, more vertically extended orbits from those in colder, rotationally supported configurations. While Milky Way–style studies analysing individual stars are not yet feasible for external galaxies, advances in instrumentation will soon make them practical; when that occurs, the choice of classification criterion for thin and thick discs will become even more consequential. Our results therefore highlight the need to state the classification criterion explicitly and to interpret it physically when analysing disc substructures and their origins.

By contrast, orbit-based dynamical modelling of IFS data (e.g. Schwarzschild orbital superposition) can already be applied to external galaxies to recover the present-day orbital distribution and to separate colder, rotationally supported material from hotter, vertically extended orbits. Ongoing efforts to incorporate chemical tagging into such models (e.g. Thater et al.\ 2022; van de Ven et al.\ 2025) point to a practical chemo-kinematic pathway that combines dynamical separation with constraints from age and metallicity. In parallel, extending high-resolution simulations that track detailed chemistry (e.g. NH2 simulation) to larger volumes will provide a stronger theoretical framework and calibration set for disc-substructure studies.

\subsection{Physical origin of the disc \msig\ relation}\label{sec:dmsig}
The existence of a tight relation between stellar mass and velocity dispersion for disc components in both the SAMI and NH samples (Figure~\ref{disc_msig}) is noteworthy given that discs are predominantly rotation-supported systems. Unlike bulges or elliptical galaxies, whose velocity dispersion reflects random motions in pressure-supported systems, discs are stabilized by ordered rotation. This raises the question of why velocity dispersion in discs correlates so strongly with stellar mass.

{Simple equilibrium arguments provide a natural baseline for} a positive disc \msig\ trend. In vertical Jeans equilibrium for an exponential stellar disc (e.g.\ van der Kruit 1988, 2002),
\begin{equation}
\sigma_z^2 \simeq \pi\,G\,\Sigma_*\,h_z,
\end{equation}
{which relates the vertical dispersion $\sigma_z$, surface density $\Sigma_*$, and scale height $h_z$. Although Eq.~(3) does not, by itself, imply a unique causal direction between $\sigma_z$ and $h_z$, combining it with empirical structural scalings with $M_*$ provides a simple expectation for how $\sigma_z$ should scale with mass, which we compare to the observed \msig\ slopes. In practice, thin and thick discs have different scale heights, so Eq.~(3) applies separately to each component, but here we use global scalings only to motivate the overall mass dependence.}

In the disc plane, Toomre stability (Toomre 1964),
\begin{equation}
Q \sim \frac{\kappa\,\sigma_R}{3.36\,G\,\Sigma_*},
\end{equation}
implies $\sigma_R \propto \Sigma_*/\kappa$ for quasi-stable discs ($Q\!\approx\!1$), where $\kappa$ is the epicyclic frequency set by the rotation curve. Systematic scalings of $\Sigma_*$ and $\kappa$ with mass therefore also drive a positive \msig\ relation. {Consistent with this expectation, we find a positive correlation (Spearman $\rho = 0.48$) between the disc stellar surface density, $\Sigma_{\ast,\rm disc} \equiv M_{\rm *,disc}/(2\pi R_{\rm e}^2)$, and the disc velocity dispersion in the SAMI spectroscopic decomposition (Figure~\ref{msig}a).}

{The observed slopes can be compared with these simple equilibrium scalings. For vertical Jeans equilibrium, $\sigma_z^2 \propto \Sigma_\ast h_z$, which implies $\sigma_z \propto M_\ast^{(1-2\alpha+\beta)/2}$ if $\Sigma_\ast \propto M_\ast/R_d^2$, $R_d \propto M_\ast^\alpha$, and $h_z \propto M_\ast^\beta$. Adopting $R_d \propto M_\ast^{0.25}$ from van der Wel et al.\ (2014) and $h_z \propto M_\ast^{0.17}$ from Tsukui et al.\ (2025) gives an expected slope of $\sim 0.33$. For a disc near constant Toomre $Q$, $\sigma_R \propto \Sigma_\ast/\kappa$; with $\kappa \propto V_c/R_d$, $R_d \propto M_\ast^{0.25}$, and adopting a stellar mass Tully-Fisher scaling $V_c \propto M_\ast^{0.3}$, this gives a slope of about 0.45. The observed disc slopes are 0.29 in the SAMI spectroscopic decomposition (Figure~\ref{msig}a), 0.34 in the SAMI circularity-based decomposition (Figure~\ref{disc_msig}a), and 0.43--0.47 in the NH simulation (Figure~\ref{disc_msig}b,c). The SAMI values lie between these two estimates, and the NH slopes are closer to the constant-$Q$ case. The disc \msig\ slopes presented here are therefore broadly consistent with these simple equilibrium expectations, although the exact slope will also depend on structural quantities such as $h_z$, $R_d$, and $\kappa$.}

Equilibrium arguments set the baseline, but the observed dispersion likely also reflects the integrated history of dynamical heating. Dynamical processes such as scattering by giant molecular clouds, bar and spiral instabilities, or minor mergers can increase random motions through secular heating and orbital diffusion (e.g.\ Fouvry et al.\ 2015, 2017), leading to elevated dispersions. More massive galaxies are likely to have experienced stronger or more sustained internal and external perturbations, leading to systematically higher levels of dynamical heating. {In addition to these mass-dependent trends, the SAMI age-spin relation shows that older stellar populations are dynamically hotter than younger populations even at fixed stellar mass and environment (Croom et al.\ 2024), consistent with heating accumulating over time. Taken together, these trends suggest that the efficiency of secular heating varies with mass, while the time available for heating modulates the dispersion within a given mass bin.} This cumulative effect may establish a mass-dependent “kinematic temperature” in discs, yielding a tight \msig\ relation consistent with the equilibrium trends discussed above.

{An alternative but compatible interpretation is that the disc \msig\ relation reflects a self-regulated state in which dynamical heating is balanced by the continued formation of young, kinematically cold stars, keeping discs close to marginal stability (Park et al.\ 2021). In this picture, orbital diffusion and other heating channels increase random motions, while ongoing star formation replenishes dynamically cold material, so that the disc maintains an approximately steady kinematic temperature set by the balance between heating and cooling. This view is consistent with the idea that discs can be reshaped by dynamical perturbations over time and that the relative contributions of disc and spheroidal components may evolve as discs settle and subsequently grow (e.g.\ Park et al.\ 2021). A related emergence perspective has been discussed for NH discs in the context of global scaling relations and marginal stability (e.g.\ Kraljic et al.\ 2024).}

Several features of our results support this overall picture. First, the disc \msig\ relation {shows a similar trend} in both SAMI and NH, and across different disc definitions (Figure~\ref{disc_msig}), suggesting that the correlation is robust. {Second, we find that the thick-to-thin dispersion ratio varies only weakly with stellar mass and galaxy type (Figure~\ref{disc_res}). Although the adopted circularity limits define the component boundaries (Figure~\ref{disc_msig_cir}), the resulting dispersion ratio also depends on the underlying circularity distribution. We find a similarly weak dependence on mass when the components are instead defined by stellar age. This weak mass dependence may suggest that vertical heating and disc self-gravity remain in broadly self-similar balance across the disc population. At the same time, the present-day constancy of this ratio does not require identical evolutionary histories, and may still allow for differences in how strongly thick discs contributed at earlier epochs before subsequent thin-disc growth.} Third, the dispersion contrast weakens when the components are defined by stellar age rather than orbital structure (Figure~6g–i). {While this is partly a consequence of defining the separation kinematically, the result suggests that an orbital based decomposition is more closely aligned with the dynamical heating state of the disc than a single global age cut.}

Overall, the disc \msig\ relation appears to encapsulate both the present-day kinematics and the integrated evolutionary history of the disc. While the detailed mechanisms may differ between thin and thick components and across morphologies, the presence of the relation in observations and simulations highlights its utility as a tracer of the secular and environmental processes that regulate disc evolution.

\subsection{Implications for the origin of thin and thick discs}
Several formation channels have been proposed for thick discs, each with distinct chemo--kinematic signatures that can be compared with our results:
\emph{(i) Early in-situ, turbulent formation.} In gas-rich, high-redshift discs, violent gravitational instability and clumpy accretion imprint a kinematically hot component at early times, producing old, $\alpha$-enhanced populations with relatively uniform ages and weak radial age gradients (e.g. Bournaud et al.\ 2009). 
\emph{(ii) External heating by satellites/minor mergers.} Repeated minor interactions heat a pre-existing disc, enhancing $\sigma_z$ and inducing flaring at large radii (e.g. Villalobos \& Helmi 2008; Qu et al.\ 2011). 
\emph{(iii) Secular in-situ heating.} Scattering by giant molecular clouds, spiral structure, and bars steadily increases random motions, building an age--velocity-dispersion relation and linking thick-disc prominence to internal disc structure (e.g.\ Aumer et al.\ 2016). Bar buckling preferentially heats the inner disc and may coexist with a box/peanut bulge (e.g.\ Fragkoudi et al.\ 2017).

In the context of these scenarios, three aspects of our findings are particularly informative. {First, the presence of a well-defined disc \msig\ trend in both SAMI and NH is consistent with a picture in which disc velocity dispersion is gradually built up through cumulative dynamical heating. However, as discussed in Section~\ref{sec:nh_offset}, the absolute level of scatter differs between SAMI and NH and should not be interpreted as direct evidence for a specific heating channel. We therefore place more emphasis on the qualitative agreement in the mean trends and on the relative behaviour of the disc subcomponents than on the absolute level of scatter.} Second, $\sigma_{\rm thick}/\sigma_{\rm thin}$ shows only a weak dependence on mass and morphology. Although this is partly influenced by the adopted circularity limits used to define the thin and thick discs (Figure~\ref{disc_msig_cir}), the result is compatible with the stability based arguments in Section~\ref{sec:dmsig}. Third, if an early, turbulent in-situ phase were the \emph{sole} origin of the thick disc, a cleaner separation by age (and often by [$\alpha$/Fe]) would be expected; instead we find that age-based selections yield a weaker dispersion contrast and substantial overlap between the age distributions of the two components, indicating that age/chemistry does not map one-to-one onto present-day dynamical thickness. This is consistent with results from NH suggesting that disc stars are typically born in highly circular orbits (Yi et al.\ 2024), while cosmological simulations such as FIRE include some cases where thick discs are largely born hot in an early turbulent phase (Yu et al.\ 2023). Whether this contrast reflects genuinely diverse formation channels or differences in numerical resolution and star formation and feedback prescriptions remains an open question.

Taken together, these trends most naturally point to a baseline of \emph{secular in-situ heating} operating over long timescales, consistent with simple vertical-equilibrium scalings. These lines of evidence do not \emph{exclude} contributions from minor mergers—occasional interactions can provide secondary, localised heating and flaring at large radii—but such events do not appear to be essential for explaining the qualitative, self-similar trends seen here. We therefore favour a mixed, chemo--kinematic picture in which internal (secular) processes establish the baseline dynamical structure, while external perturbations provide incremental heating.

\section{Summary and Conclusions}
We have analysed disc kinematics in the SAMI Galaxy Survey and in the NewHorizon simulation, comparing circularity-based and age-based definitions of thin and thick discs. We measured the relation between stellar mass and velocity dispersion for disc components, quantified the thin and thick disc contrast, and assessed how classification scheme affects inferred trends. Our key findings are as follows.

\begin{itemize}
\item {Tight disc mass–dispersion relation:} both SAMI and NH show a well-defined disc \msig\ relation, nearly parallel to the bulge relation, highlighting the physical relevance of velocity dispersion in rotation-supported discs (Figure~\ref{disc_msig}). We find method-dependent offsets (Section~\ref{sec:sch_offset}): \emph{Schwarzschild} modelling yields slightly higher $\sigma_{\rm e}$ than spectroscopic bulge–disc decompositions. We also find an observation–simulation offset (Section~\ref{sec:nh_offset}): at fixed $M_*$, NH has lower $\sigma_{\rm e}$ than SAMI, especially at low mass, consistent with known simulation–observation differences.

\item {Weak mass dependence of the thin--thick contrast:} {the thick component is systematically hotter than the thin component, and the thick-to-thin dispersion ratio varies only weakly with stellar mass, bulge fraction, and thick-disc light fraction for both circularity-based and age-based selections (Figure~\ref{disc_res}). This behaviour remains robust against reasonable variations in the adopted circularity and age thresholds (Sections~\ref{circularity_selection} and \ref{age_selection}).}

\item {Age cuts do not reproduce the circularity-based split:} {selecting disc components by stellar age yields a weaker dispersion contrast than the circularity-based decomposition (Figure~\ref{disc_res}), as expected because the latter is defined directly in kinematic space. Consistent with this, the circularity-selected components show broad and overlapping age distributions (Figure~\ref{disc_age}), and no single global age threshold (4--8~Gyr tested) reproduces the circularity-based split across the mass range (Figure~\ref{disc_msig_age}).}
\end{itemize}

Taken together, these results are consistent with a baseline set by vertical equilibrium scalings, with cumulative secular heating building a mass-dependent kinematic temperature over time. A merger-dominated picture is not required to explain the global, self-similar trends, although occasional minor interactions may contribute localised heating and flaring. A purely early, turbulent origin acting alone would also predict a cleaner separation by age than observed. The present analysis is limited by sample size (and associated selection and measurement uncertainties), so the conclusions should be regarded as provisional.

Looking ahead, chemo–kinematic dynamical modelling of IFS data offers a practical route to combine orbit families with age and abundance constraints. Targeted tests of flaring and the radial behaviour of $\sigma_z/\sigma_R$, together with higher resolution, chemistry-tracking simulations over larger volumes, will improve calibration for disc-substructure diagnostics and help clarify the contribution of mergers.

\section*{Acknowledgements}
The SAMI Galaxy Survey is based on observations made at the Anglo-Australian Telescope. The SAMI was developed jointly by the University of Sydney and the Australian Astronomical Observatory. The SAMI input catalogue is based on data taken from the SDSS, the GAMA Survey, and the VST ATLAS Survey. The SAMI Galaxy Survey is supported by the Australian Research Council ASTRO 3D, through project number CE170100013, the Australian Research Council Centre of Excellence for All-sky Astrophysics (CAASTRO), through project number CE110001020, and other participating institutions. SO acknowledges support from the Korean National Research Foundation (NRF) (RS-2023-00214057), as well as ongoing support from DL. SKY acknowledges support from the Korean National Research Foundation (RS-2025-00514475 and RS-2022-NR070872). This work was granted access to the HPC resources of KISTI under the allocations KSC-2021-CRE-0486, KSC-2022-CRE-0088, KSC-2022-CRE-0344, KSC-2022-CRE-0409, KSC-2023-CRE-0343, KSC-2024-CHA-0009, and KSC-2025-CRE-0031 and of GENCI under the allocation A0150414625 and A0180416216. The large data transfer was supported by KREONET which is managed and operated by KISTI. This work was partially supported by the Institut de Physique des deux infinis of Sorbonne Université and by the ANR grant ANR-19-CE31-0017 of the French Agence Nationale de la Recherche. CP's research was in part supported by the Segal ANR-19-CE31-0017 (http://secular-evolution.org) and ANR-25-CE31-4684 (https://www.secular-bars.org/). MLPG acknowledges support from the ARC grant DP190102714. This project has received financial support from the CNRS through the MITI interdisciplinary programs.

\section*{Data Availability}
The observational data used in this study are publicly available through Astronomical Optics' Data Central service at https://datacentral.org.au/ as part of the SAMI Galaxy Survey Data Release 3. The simulation data used in this study were obtained through collaboration with the NewHorizon team.

\label{lastpage}
\end{document}